\documentclass[aps,prx,twocolumn,amsmath,amssymb,showpacs,superscriptaddress,notitlepage]{revtex4-1}
\usepackage{amsmath,amssymb,amsfonts,bm}
\usepackage{graphicx}
\usepackage{dcolumn}
\usepackage{mathrsfs}
\usepackage{bbold}
\usepackage{dsfont}
\usepackage{dcolumn}
\usepackage[colorlinks=true,linkcolor=blue,citecolor=blue, urlcolor=blue,bookmarks=false]{hyperref}
\usepackage{changes}

\begin{document}

\title{Tunable corner-like modes in generalized quadrupole topological insulator}

\author{Rui Chen}
\affiliation{Department of Physics, Hubei University, Wuhan 430062, China}
\author{Bin Zhou}
\affiliation{Department of Physics, Hubei University, Wuhan 430062, China}

\author{Dong-Hui Xu}\email[]{donghuixu@cqu.edu.cn}
\affiliation{Department of Physics and Chongqing Key Laboratory for Strongly Coupled Physics, Chongqing University, Chongqing 400044, China}
\affiliation{Center of Quantum Materials and Devices, Chongqing University, Chongqing 400044, China}

\begin{abstract}

Higher-order topological insulators harbor unique corner modes that hold immense potential for applications in information storage. However, the practical manipulation of these states has been constrained by the fixed positions and energies of conventional corner modes. In this work, we present a theoretical framework for generating topologically protected corner-like modes in higher-order topological insulators, exhibiting unprecedented tunability in their positions. These corner-like modes are characterized by a quantized generalized quadrupole moment, indicative of the presence of fractional charges. Moreover, we demonstrate the remarkable ability to modulate the energy of these topological corner modes. Our findings pave the way for the controlled manipulation of corner modes in higher-order topological insulators, opening up new avenues for their applications in advanced information technologies.
\end{abstract}
\maketitle
{\color{blue}\emph{Introduction}.}---Higher-order topological insulators (HOTIs) have captivated the research landscape in recent years, distinguished by their capacity to host topologically protected gapless boundary modes in lower dimensions than their bulk~\cite{Schindler2018SciAdv,Langbehn17PRL,
Benalcazar17PRB,Song17PRL,Yang2024JPCM,Fan2021FP,ZengC2019PRL,Benalcazar2017Science}. The quadrupole topological insulator, a prime example of a HOTI, exhibits unique properties, including a quantized quadrupole moment in its bulk and the resulting localized states at its zero-dimensional geometric corners~\cite{Benalcazar2017Science,ChenR20PRL,LiCA20PRL,Varjas19PRL,Hua20PRB,Mao24PRB}. Its successful realization across various experimental platforms, ranging from acoustics to electrical circuits~\cite{Xue2018NatMat,Qi2020PRL,Xue19PRL,LiX2021FP,Kim2020LSA,Schulz2022NC,Mittal2019NatPho,SerraGarcia2018Nature,Kempkes2019NatMat,Peterson2018Nature,Lv2021ComPhys}, highlights its versatility and potential for diverse applications. These corner modes, shielded by their topological nature, hold promise for information storage, and even fault-tolerant quantum computing~\cite{Wu2020PRLnonabelian}.

While past research has predominantly focused on corner modes fixed in both spatial location (geometric corners) and energy level, the potential for their manipulation remains largely unexplored. In this study, we aim to address this gap by introducing a novel approach to generate and manipulate corner-like modes in a higher-order topological insulator. By introducing a spin-orbital potential, we demonstrate the emergence of corner-like modes localized at the intersection of two potential domains, rather than the conventional geometric corners. This discovery unlocks new avenues for controlling the spatial distribution of these modes, as we further demonstrate their position can be tuned by adjusting the spin-orbital potential strength. We also unveil the topological nature of these corner-like modes, characterized by a quantized generalized quadrupole moment. Beyond spatial control, we investigate the manipulation of corner mode energy. By introducing a periodic spin-orbital potential, we can tune these energies, adding another dimension of control and flexibility to HOTI manipulation.

\begin{figure}[thtb]
\centering
\includegraphics[width=\columnwidth]{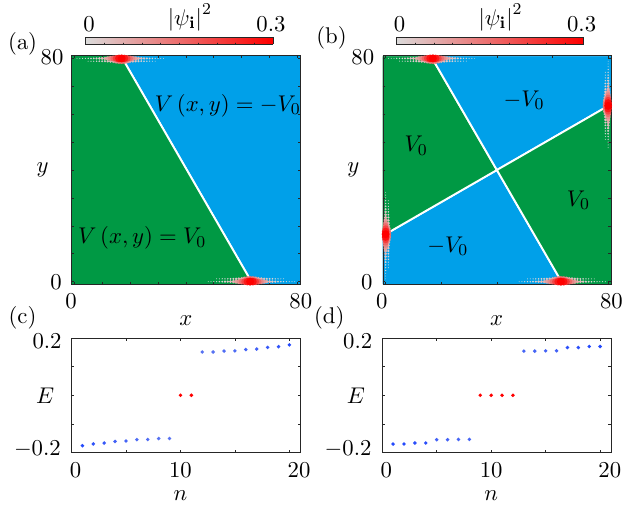}
\caption{[(a)-(b)] Schematic illustration of $V\left(x,y\right)$. The green and cyan regions denote $V\left(x,y\right)=V_0$ and $V\left(x,y\right)=-V_0$, respectively. The red points, corresponding to the probability distribution $\left|\psi_{\mathbf{i}}\right|^2$ of the zero energy states in panels (c)-(d), denote the corner-like modes. (c) and (d) show the energy spectra of the system in (a) and (b), respectively. The red and blue points correspond to the corner-like states and edge states, respectively. }
\label{fig:mass}
\end{figure}

Our findings not only deepen the understanding of higher-order topological insulators but also pave the way for the controlled manipulation of corner states in HOTIs, unlocking new possibilities for their application in advanced information technologies.

{\color{blue}\emph{Model}.}---To frame our discussion, we consider a first-order topological insulator described by the Bernevig-Hughes-Zhang (BHZ) Hamiltonian~\cite{Bernevig2006Science} with an additional spin-orbital potential
\begin{equation}
H=H_\text{QSH}\left(k_x,k_y\right)+H^\prime(x,y),
\label{Eq_Hamiltonian}
\end{equation}
where $H_\text{QSH}=(M-Bk^2)\sigma_0\tau_z+A k_x \sigma_z\tau_x+A k_y \sigma_0\tau_y$ is the four band BHZ Hamiltonian in momentum space with $k^2=k_x^2+k_y^2$.  $\sigma$ and $\tau$ are Pauli matrices representing spin and orbital degrees of freedom, respectively.  $H_\text{QSH}$ supports a pair of gapless helical edge states within the bulk gap. $H^\prime(x,y)=V(x,y)\sigma_x\tau_x$ represents a real-space spin-orbital potential term with its spatial distribution determined by $V(x,y)$. In the following, we will show that $H^\prime(x,y)$ can gap out the helical edge states by breaking time-reversal symmetry, leading to the emergence of corner-like modes at the interface between domains of opposite mass. For our numerical calculations, we discretize the effective Hamiltonian on a two-dimensional square lattice with lattice constant $a=1$. The parameters are taken as $A=1$, $B=1$, and $M=1$.

{\color{blue}\emph{Corner-like modes}.}---We investigate a system with a nonuniform potential distribution, $V(x,y)$, as shown in Fig.~\ref{fig:mass}(a), consisting of two domains with opposite potentials $\pm V_0$. This configuration gives rise to a unique energy spectrum in Fig.~\ref{fig:mass}(c), featuring two zero-energy modes located inside the gapped edge states. The probability distributions of these zero-energy modes depicted in Fig.~\ref{fig:mass}(a) reveal their localized nature at the corners where the two domains intersect. We dub the localized states as corner-like modes.

A notable feature of these corner-like modes is their tunability. By manipulating the potential distribution $V(x,y)$, we can control the position and number of corner-like modes,  as demonstrated in  Figs.~\ref{fig:mass}(b) and \ref{fig:mass}(d). This flexibility distinguishes them from conventional corner modes found in quadrupole topological insulators. In the latter case, the probability distribution is confined to the geometric corners and cannot be repositioned without altering the geometry itself.

{\color{blue}\emph{Orientation-independent Dirac mass of the edge states}.}---To gain an intuitive understanding of the emergent corner-like modes, let us consider the simplest case, i.e., a uniformly distributed $V\left(x,y\right)=V_0$. This maintains translational symmetry, allowing us to analyze the system using momentum space within the framework of edge theory~\cite{Shen2017TI,LuHZ2010PRB,YanZB18PRL,Agarwala2020}. Treating $V_0$ as a perturbation, we focus on the states residing on the edge with a polar angle $\beta$. Through the coordinate transformation, $k_{x} =k_{x}^{\prime }\cos \beta +k_{y}^{\prime }\sin \beta $,
$k_{y} =k_{y}^{\prime }\cos \beta -k_{x}^{\prime }\sin \beta$, and replacing $k_{y}^{\prime }\rightarrow -i\partial_{y}^{\prime }$, we obtain the wave function of the edge states on the $y^\prime=0$ surface of $H_\text{QSH}$. Projecting the total Hamiltonian $H$ on the wave function at $k_{x}^{\prime }=0$ leads to an effective model for edge states (see Sec.~SI A in Supplemental Material~\cite{supp})
\begin{equation}
H_\text{edge}(k_{x}^{\prime})=\begin{pmatrix}
Ak_{x}^{\prime} & V_{0}e^{i\beta}\\
V_{0}e^{-i\beta} & -Ak_{x}^{\prime}%
\end{pmatrix}.
\end{equation}
The eigenvalues of the above effective Hamiltonian are given by $E_{\text{edge}}^{\pm}(k_{x}^{\prime})$$=$$\pm \sqrt{A^2 k_x^{\prime2}+V_0^2}$, which describes a massive Dirac fermion with an orientation-independent topological mass $V_0$. Note that the mass of the edge state does not depend on the angle $\beta$. The localized state at the corner between two regions with opposite $V_0$ is a manifestation of the Jackiw-Rebbi mechanism~\cite{Jackiw76PRD}. This mechanism predicts the emergence of a zero-energy mode at the interface (in this case, the corner) between regions of opposite mass. It's important to highlight that these corner-like modes are distinct from conventional corner modes found in other topological systems~(see Refs.~\cite{YanZB18PRL,Agarwala2020} and Sec.~SI B in Supplemental Material~\cite{supp} for more details).  Conventional corner modes typically exhibit an orientation-dependent topological mass, meaning their properties change with the angle.  In contrast, the corner-like modes discussed here have an orientation-independent mass, making them a unique feature of this
higher-order topological insulator.

\begin{figure}[t]
\centering
\includegraphics[width=\columnwidth]{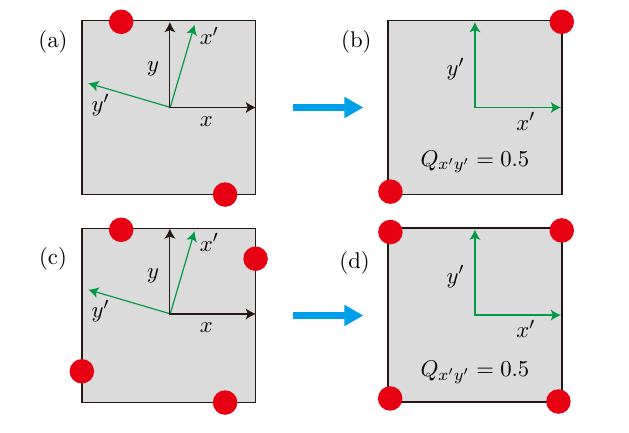}
\caption{Schematic illustration for calculating the generalized quadrupole moment. The red points in (a) and (c) label the positions of the corner-like modes. [(b) and (d)] After a proper coordinate transformation, the positions of the corner-like modes are transformed to the corners (see Sec.~SII in Supplemental Material~\cite{supp} for more details).}
\label{fig:mass_illustration}
\end{figure}

{\color{blue}\emph{Generalized quadrupole moment}.}---Corner-like modes have a topological origin, characterized by a quantized generalized quadrupole moment. The general quadrupole moment is defined as~\cite{Wheeler19PRB,Kang2019PRB,LiCA20PRL}
\begin{equation}
Q_{x y}=\left[\frac{1}{2 \pi} \operatorname{Im} \ln \operatorname{det}\left(U_o^{\dagger} \hat{D} U_o\right)-Q_0\right] \bmod 1,
\end{equation}
where $U_o=\left(\left|\psi_1\right\rangle,\left|\psi_2\right\rangle, \ldots,\left|\psi_{N_{\text {occ }}}\right\rangle\right)$ is a matrix consisting of occupied eigenstates of the first-quantization Hamiltonian under periodic boundary conditions, $\hat{D}=\exp{\left(2 \pi \hat{x} \hat{y} / L^2\right)}$ with $\hat{x}$ and $\hat{y}$ being the coordinate operator, and $Q_0$ is the contribution from the background positive charge distribution.

In the original real-space coordinates $(x,y)$, the system depicted in Fig.~\ref{fig:mass_illustration}(a) exhibits a trivial quadrupole moment $Q_{xy}=0$.  However, by using a proper coordinate transformation
$(x,y)\rightarrow (x^\prime,y^\prime)$, the corner-like modes can be shifted to the corners, as shown in Fig.~\ref{fig:mass_illustration}(b). This transformed system is then characterized by a generalized quantized quadrupole moment $Q_{x^\prime y^\prime}=0.5$~\cite{Tao2023SciPhy} in the new coordinates. Moreover, we find that the system with a greater number of corner-like modes, like the one in Fig.~\ref{fig:mass_illustration}(c), can also be characterized by a quantized generalized quadrupole moment after a proper coordinate transformation, as demonstrated in Fig.~\ref{fig:mass_illustration}(d).

\begin{figure}[tpb]
\centering
\includegraphics[width=\columnwidth]{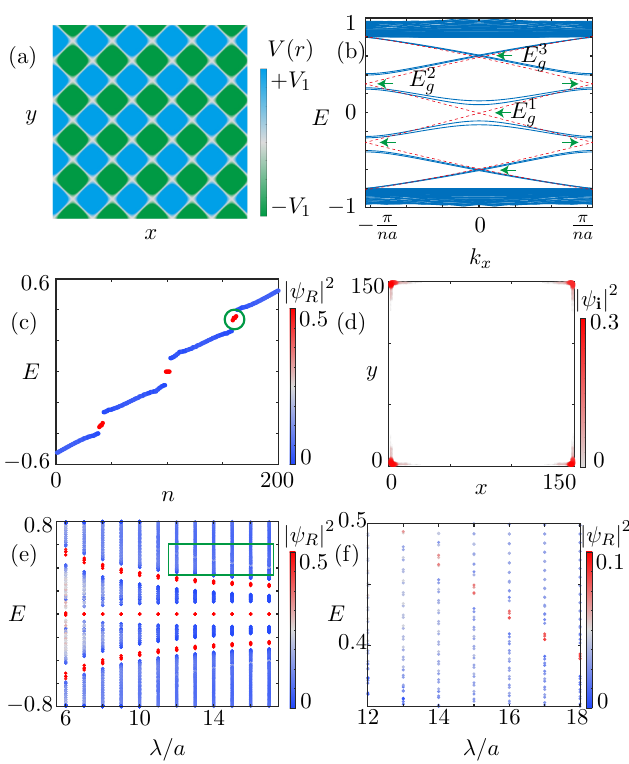}
\caption{(a) Schematic illustration of the periodically distributed $V(x,y)$. (b) Energy spectra of the system with open boundary along the $y$-direction and periodic boundary condition along the $x$-direction. In the weak field limit of the periodic field ($V_{1}\rightarrow 0$), the Brillouin zone is folded and induces more band crossings labeled by the green arrows. For nonzero field strength ($V_{1}=0.15$), the band crossings are gapped out.
(c) Energy spectra of the system with open boundary along both the $x$- and $y$-directions. The red states are the corner states shown in (d), and the blue states are the edge states.   The strip length is $L_y=40 a$ in (b). The system size is $n_x=n_y=150 a$ in (c)-(d), and $n_x=n_y=10 \lambda$ with the wavelength $\lambda=10 a$ in (e)-(f). In all panels, we take $\theta=\pi$ and $\left|\psi_R\right|^2=\sum_{\mathbf{i}\in\text{Corners}}\left|\psi_{\mathbf{i}}\right|^2$.}
\label{fig:spectrum}
\end{figure}

{\color{blue}\emph{Periodic field modulated corner modes}.}---Previous research has shown that the energy of corner modes is typically fixed~\cite{Schindler2018SciAdv,Langbehn17PRL,
Benalcazar17PRB,Song17PRL,Yang2024JPCM,Fan2021FP,Benalcazar2017Science}. In this section, we show that the energy of the corner modes can be tuned by a periodic spin-orbital potential, as illustrated in Fig.~\ref{fig:spectrum}(a). The periodic potential takes the form
\begin{equation}
V(x,y)=V_1\left[\cos\left(q x\right)+\cos\left(q y+\theta\right)\right],
\end{equation}
where $V_1$ is the amplitude of the periodic term, $\theta$ is the phase, and $q=2\pi/\lambda$ with $\lambda$ the wavelength.

Figure~\ref{fig:spectrum}(b) shows the energy spectra of the system under open boundary condition along the $y$-direction and periodic boundary condition along the $x$-direction. The system preserves the supercell transitional symmetry and can be investigated in a folded Brillouin zone, leading to the additional band crossings in the weak field limit [see the red dashed lines with $V_{1}\rightarrow 0$ in Fig.~\ref{fig:spectrum}(b)]. For nonzero field strength with $V_{1}=0.15$, the band crossings are gapped out due to spin mixing [the blue sold lines in Fig.~\ref{fig:spectrum}(b)]. It is noticed that $E_g^{1,2}$ (i.e., the magnitudes of the energy gaps at lower energies) is much larger compared to $E_g^3$ (i.e., the magnitude of the energy gap at higher energies).

When open boundary conditions are applied in both $x$- and $y$-directions (Fig.~\ref{fig:spectrum}(c)), three distinct energy gaps emerge, each hosting four in-gap corner modes. Figure~\ref{fig:spectrum}(d) shows the probability of the four in-gap modes enclosed by the green circle shown in Fig.~\ref{fig:spectrum}(c) as an example, which indicates that the in-gap modes are corner modes. Moreover, we find that the bulk of the system is characterized by a quantized quadrupole moment $Q_{xy}=0.5$, which reveals the topological nature of the corner modes in Figs.~\ref{fig:spectrum}(c) and \ref{fig:spectrum}(d).

Figure~\ref{fig:spectrum}(e) shows the energy spectra of the periodic-field-modulated QSH system as a function of the wave length $\lambda$ with the system size $L_x=L_y=10 \lambda$. The zero-energy corner modes remain stable with the increasing $\lambda$. While, the energy of the corner modes at higher energies decreases with the increasing $\lambda$.  Thus, the periodic spin-mixing field provides a method to control the energy of the corner modes. In addition, at higher energies, there also exists a tiny energy gap hosting the corner modes [see the green rectangle in Fig.~\ref{fig:spectrum}(e) and Fig.~\ref{fig:spectrum}(f)]. This tiny energy gap originates from the tiny energy gap $E_g^3$ in Fig.~\ref{fig:spectrum}(b).


{\color{blue}\emph{Experimental realization}.}---The above theoretical scenarios can be realized in solid materials. For example, the orientation-independent Dirac mass of the edge states can be induced in the two coupled HgTe quantum wells with $s$-wave interlayer excitonic pairings~\cite{Budich2014PRL,LiuZR21PRBL}. Specifically, the low-energy Hamiltonian of the exciton condensate reads
\begin{equation}
\begin{aligned}
H(\mathbf{k})= & A\left(k_x \sigma_x s_z+k_y \sigma_y\right)+\left[M-B\left(k_x^2+k_y^2\right)\right] \sigma_z \\
& +\Delta_X(x,y) \tau_y \sigma_x s_x,
\end{aligned}
\end{equation}
where $s$, $\sigma$, and $\tau$ are Pauli matrices acting on spin, orbital, and layers degrees of freedom, respectively. Here, $\Delta_X(x,y)$ is the pairing strength and can be tuned by an external gate voltage. In a local region with $\Delta_X(x,y)=\Delta_0$, the helical edge state is gapped out with an orientation-independent Dirac mass $\Delta$~\cite{LiuZR21PRBL}. Thus, the corner-like modes can be realized by proper engineering of the gate voltage. On the other hand, the scenario can be experimentally realized in electric circuits, considering the Dirac Hamiltonian had been experimentally realized in a recent work~\cite{Yang2023Comphys}. Therefore, these features offer
the possibility of realizing our proposal in the future experiments.

{\color{blue}\emph{Conclusion}.}---In this study, we have proposed a theoretical framework for generating and manipulating corner-like modes in higher-order topological insulators. By utilizing a spin-orbital potential, we have demonstrated the emergence of tunable corner-like modes with a quantized generalized quadrupole moment. We have also established the ability to modulate the energy of these modes through the application of a periodic spin-orbital potential. Our findings offer new insights into the manipulation of corner modes in HOTIs and pave the way for their exploitation in advanced technological applications.

\begin{acknowledgments}
D.-H.X. was supported by the NSFC (under Grant Nos.~12074108 and 12347101), the Natural Science Foundation of Chongqing (Grant No.~CSTB2022NSCQ-MSX0568). R.C. acknowledges the support of the NSFC (under Grant No. 12304195) and the Chutian Scholars Program in Hubei Province.  B.Z. was supported by the NSFC (under Grant No. 12074107), the program of outstanding young and middle-aged scientific and technological innovation team of colleges and universities in Hubei Province (under Grant No. T2020001) and the innovation group project of the natural science foundation of Hubei Province of China (under Grant No. 2022CFA012).
\end{acknowledgments}

%
%
%
\bibliographystyle{apsrev4-1-etal-title_6authors}
\bibliography{refs-transport,refs-transport_v1}

\end{document}